\documentclass[12pt,aps,nofootinbib,preprintnumbers,groupedaddress,letterpaper]{article}

\usepackage{hyperref}

\usepackage{cite}

\usepackage[dvipsnames]{xcolor}
\usepackage[normalem]{ulem}
\usepackage{amsmath}
\usepackage{enumerate}
\usepackage{mathtools}

\usepackage{amsfonts}
\usepackage{amssymb}

\usepackage[colorinlistoftodos,prependcaption,textsize=tiny]{todonotes}

\usepackage{graphicx}

\usepackage{enumitem}
\usepackage{pifont}
\def\teller{P\"{o}schl-Teller }

\def\psyk{\mathsf{p}}

\newcommand\comment[1]{}

\newcommand\thooft{'t~Hooft }

\newcommand\schr{Schr\" odinger }

\newcommand\ov{\over }

\def\le{\left}
\def\ri{\right}

\def\({\left(}
\def\){\right)}
\def\<{\langle}
\def\>{\rangle}

\newcommand\half{{\ensuremath{\frac{1}{2}}}}
\newcommand\p{\ensuremath{\partial}}

\newcommand\field[1]{{\ensuremath{\mathbb{{#1}}}}}

\newcommand{\CC}{\field{C}}

\newcommand{\RR}{\field{R}}

\newcommand{\ZZ}{\field{Z}}

\newcommand{\be}{\begin{equation}}
\newcommand{\ee}{\end{equation}}
\newcommand{\bea}{\begin{eqnarray}}
\newcommand{\eea}{\end{eqnarray}}
\newcommand{\bwt}{\begin{widetext}}
\newcommand{\ewt}{\end{widetext}}

\newcommand{\bi}{\begin{itemize}}
\newcommand{\ei}{\end{itemize}}
\newcommand{\ben}{\begin{enumerate}}
\newcommand{\een}{\end{enumerate}}
\newcommand{\bca}{\begin{cases}}
\newcommand{\eca}{\end{cases}}
\newcommand{\bln}{\begin{align}}
\newcommand{\eln}{\end{align}}
\newcommand{\bst}{\begin{split}}
\newcommand{\est}{\end{split}}

\renewcommand{\Im}{\textrm{Im}\,}

\def\Xint#1{\mathchoice
{\XXint\displaystyle\textstyle{#1}}%
{\XXint\textstyle\scriptstyle{#1}}%
{\XXint\scriptstyle\scriptscriptstyle{#1}}%
{\XXint\scriptscriptstyle\scriptscriptstyle{#1}}%
\!\int}
\def\XXint#1#2#3{{\setbox0=\hbox{$#1{#2#3}{\int}$}
\vcenter{\hbox{$#2#3$}}\kern-.5\wd0}}

\def\dashint{\Xint-}

\begin{document}

\begin{titlepage}

\begin{flushright}
QMUL-PH-24-13\\
\end{flushright}

\vspace{5mm}

  \begin{center}

\centerline{\Large \bf {Quantizing the folded string in AdS$_2$}}

\bigskip
\bigskip

{\bf David Vegh}

\bigskip

\small{
{ \it   Centre for Theoretical Physics, Department of Physics and Astronomy \\
Queen Mary University of London, 327 Mile End Road, London E1 4NS, UK}}

\medskip

{\it email:} \texttt{d.vegh@qmul.ac.uk}

\medskip

{\it \today}

\bigskip


\begin{abstract}

In two-dimensional flat space, the oscillatory motion of a closed folded string---or alternatively, two massless particles connected by a string---can be quantized using the 't Hooft equation.
This paper presents an alternative method for quantizing the folded string in anti-de Sitter space.
By using variables inspired by integrability, setting $g \equiv  {(R_\text{AdS})^2 \ov 2\pi \alpha'} $ to a specific $\psyk$-dependent $\mathcal{O}(1)$ value, and applying a particular boundary condition to the antisymmetrized wavefunction, we obtain a spectrum that precisely matches that of fermion bilinear operators in the disorder-averaged Sachdev-Ye-Kitaev model with $\psyk$-fermion interactions.

\end{abstract}

\end{center}

\end{titlepage}

\vskip-1.5cm
\tableofcontents

 \clearpage

\section{Introduction}

The quantization of physical systems is a non-trivial step that requires careful consideration and can often be performed in inequivalent ways, depending on the  choice of canonical variables. A particularly interesting topic is the quantization of extended objects, such as the relativistic string, which is one of the building blocks of string theory.
In the simplest case, one considers a low-dimensional target space, which restricts the possible motion of the string. In two dimensions, there is no transverse direction to the string, so its oscillations can only be longitudinal. Significant  work has been done in this area since the early days of the dual models for strong interactions, see \cite{Patrascioiu:1974un, PhysRevD.13.2364, Bars:1975dd, BARS1976413, Bardeen:1976yt, Artru:1979ye,  Bars:1994sv}.
For recent studies  on folded strings, refer to \cite{Ficnar:2013wba, Donahue:2019adv, Donahue:2019fgn, Donahue:2022jxu}, and for their higher-dimensional counterparts, known as segmented strings, see \cite{Vegh:2015ska, Callebaut:2015fsa, Vegh:2016hwq, Gubser:2016wno, Gubser:2016zyw, Vegh:2016fcm, Vegh:2021jhl, Vegh:2021jqo, Vegh:2023snc}. These discrete strings have a finite number of degrees of freedom and one might hope that their quantization is easier to address. For other works on discrete strings, see \cite{Giles1977, Klebanov:1988ba}, and for a recent discretization, see the fishchain model \cite{Gromov:2019aku, Gromov:2019bsj}.

\begin{figure}[h]
\begin{center}
\includegraphics[width=6cm]{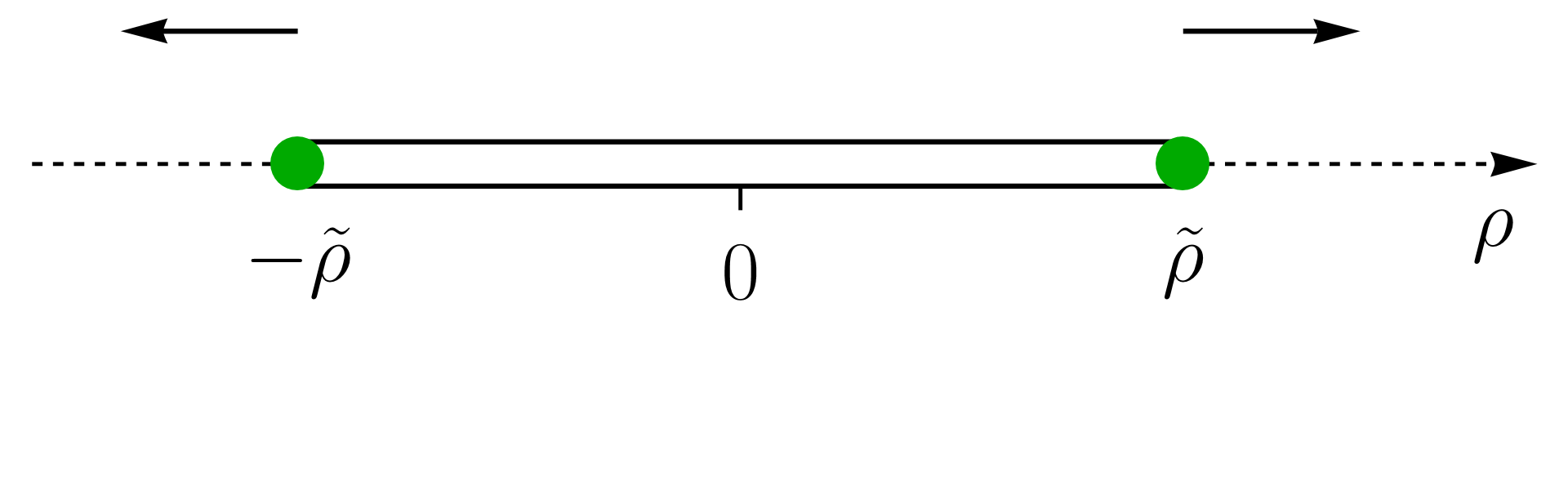} \qquad\qquad
\includegraphics[width=2.5cm]{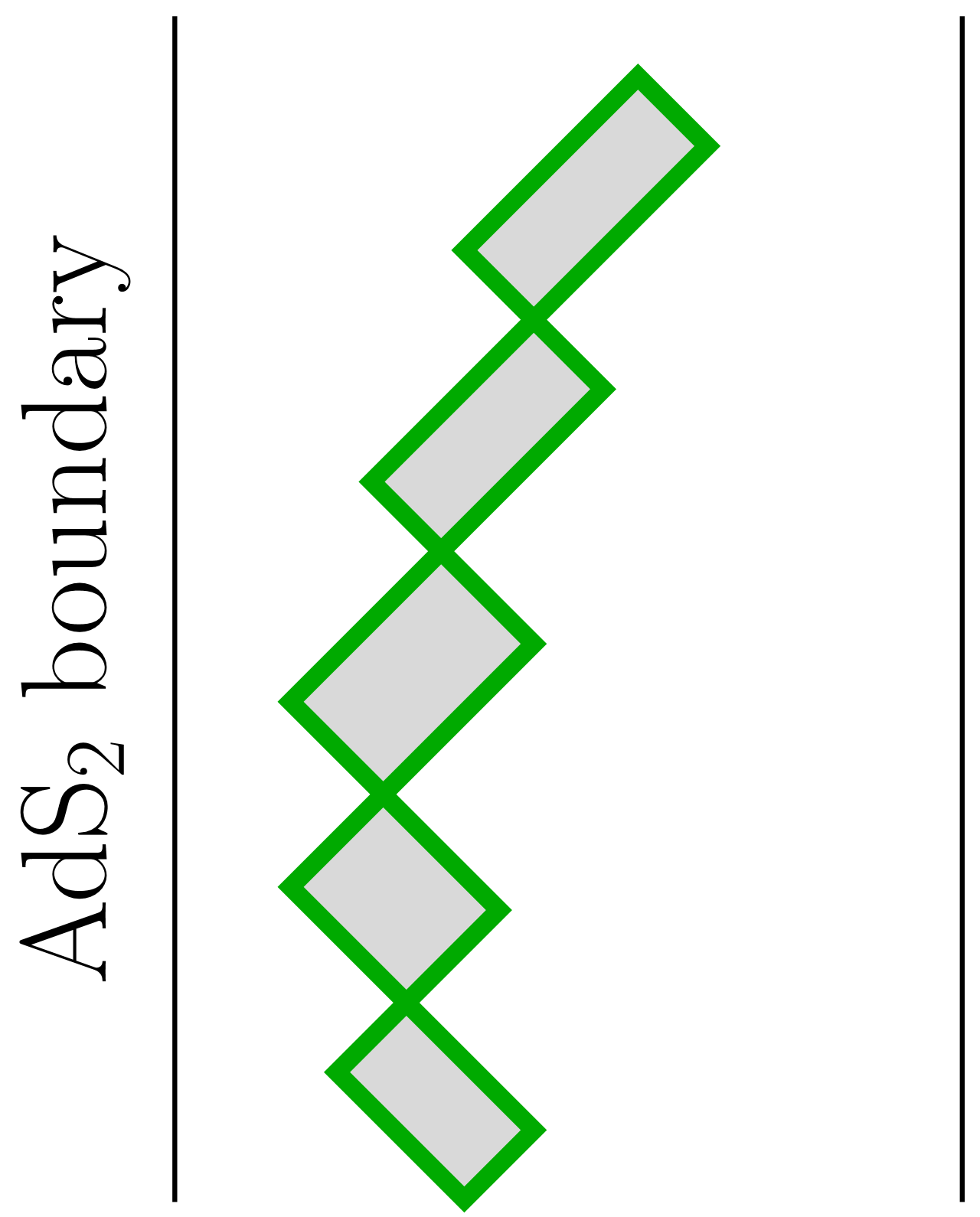}
\caption{\label{fig:yoyo}
{\it Left:}
A folded  string consists of two massless particles (green dots) connected by two string segments.  In the center-of-mass frame the position and momentum of one of the particles is denoted by $\tilde \rho$ and $\tilde p$, respectively.
{\it Right:} Folded string in AdS$_2$ space.
The particles move with the speed of light and suddenly change direction (see zigzagging  worldlines). }
\end{center}
\end{figure}

The aim of this paper is to reexamine the quantization of the simplest yo-yo string, consisting of two massless particles connected by a string. An almost equivalent system is the closed folded string, where the string in the middle is doubled (see Figure \ref{fig:yoyo}, left). Since the string dynamics on maximally symmetric target spaces is classically integrable \cite{Pohlmeyer:1975nb}, this configuration becomes particularly useful when discussing spectral curves, which require the string to be closed.
Spectral curves are algebraic curves that encode the conserved quantities in integrable systems. In the context of classical strings, they provide a powerful tool for solving the equations of motion and can also be used to analyze the spectrum of physical states in the quantum theory. Although we describe a system with a single degree of freedom (in the center-of-mass frame), techniques and canonical variables borrowed from integrability offer a new approach to quantization.

In the full string theory, the (open or closed) string can have an infinite number of longitudinal degrees of freedom, which means that it can have an arbitrary number of folds. Our plan here is to quantize the theory restricted to the two-fold sector, with the expectation that the resulting spectra will have relevance for the full theory. Notably, classical solutions with $n$ folds can be embedded within  the theory of the $n'$-fold string if $n'\geq n$. If the full theory can be quantized such that it is quantum integrable, there is hope that the two-fold spectrum is embedded within the full spectrum, as integrable theories tend to have a stronger connection to their classical counterparts compared to ordinary field theories.

The string exerts a constant force on the particles at its endpoints, resulting in oscillatory motion. Since the particles are massless, they follow null-geodesics in the target space. They carry finite momentum, and their velocities experience sudden jumps when the momentum changes sign (see Figure \ref{fig:yoyo}, right).
In two-dimensional flat space, one possible quantization of the yo-yo string employs the \thooft equation \cite{tHooft:1974pnl}, which also describes the spectrum of mesons in 2d large-$N$ QCD, where the quarks are in the fundamental representation. The massless \thooft equation can be derived quickly by considering the effective Hamiltonian for the string
\be
  \label{eq:hketto}
  H_2 = |p_1|+|p_2| + \kappa |x_1-x_2| \, ,
\ee
where $\kappa \sim (\alpha')^{-1}$ is the string tension, and $x_{1,2}$ are the coordinates of the string endpoints. The last term is a linear potential representing the string's pull.

Let us switch to center of mass and relative coordinates,
\be
  \nonumber
  x_{1,2} = x_0 \pm {x' \ov 2} \, ,
  \qquad  p_{1,2} =  {p_0\ov 2} \pm p' \, .
\ee
In the infinite momentum frame ($p_0 \to \infty$), the string's rest mass $M$ can be expressed as
\be
  \label{eq:infmom}
  M^2 = H_2^2 - p_0^2 \approx 2p_0(H_2-p_0) =  2 \kappa|s| \, ,
\ee
where $s := x'(p_1 + p_2)$ is a signed action variable. The conjugate momentum fraction variable is given by
\be
  \label{eq:momfrac}
  z := {p_1 \ov p_1 + p_2} \, ,   \qquad \{ s,z \} =   1 \, .
\ee
For a consistent description, one must project on positive momentum modes \cite{Lenz:1995tj}, restricting the domain of $z$ to the interval $[0,1]$. By considering wavefunctions in the $z$-basis and promoting $s$ to an operator acting on them, one obtains the \thooft equation with zero renormalized quark masses,
\be
  \label{eq:thooft}
   \mu^2 \varphi(z) =  - \dashint_{0}^1 dz'  {\varphi(z') \ov (z' - z)^2} \, ,
\ee
where we defined $\mu^2 = {\pi M^2 \ov 2\kappa}$.
The resulting spectrum is discrete and it is close to a linear Regge trajectory: $\mu^2 \approx \pi^2(n+{3\ov 4})$, $n=0,1,\ldots$. Exact solutions are not known, and calculations must be performed either numerically \cite{tHooft:1974pnl, HANSON1977477, Brower:1978wm} or via an expansion \cite{Fateev:2009jf} (see also \cite{Ambrosino:2023dik} for recent results). The latter method begins by inverting the \thooft equation to derive a   finite difference equation \cite{Fateev:2009jf}. Let us first switch to a new coordinate $p$ defined by
\be
  \label{eq:pdef}
  p := \log{z \ov 1-z} \, .
\ee
We get
\be
  \nonumber
   \mu^2 {\varphi(p)}  =
   - \cosh^2 \le({p\ov 2}\ri) \, \dashint_{-\infty}^{\infty} dp'
   {\varphi(p') \ov \sinh^2\le({p-p'\ov 2}\ri)} \, .
\ee
The canonical conjugate of $p$ will be denoted by $\nu$ in the flat space case and by $u$ in the AdS case, respectively. These pairs play a central role in the paper as they parametrize the spectral curve. After Fourier transforming $ \varphi(p) \to \psi(\nu)$, we get
\be
   \nonumber
 {\mu^2 \ov  2\pi} \int_{-\infty}^{\infty} d\nu' S(\nu - \nu') \psi(\nu') = \pi \nu \coth\le({\pi \nu  } \ri) \psi(\nu)  \, ,
\ee
with the kernel $ S(\nu) = {\pi \nu \ov   \sinh\le({\pi \nu } \ri) } $. This integral equation can be shown to be equivalent to the finite difference equation
\be
  \label{eq:fdifeq}
   Q(\nu+i)+Q(\nu-i) -2 Q(\nu) = -{\mu^2 } { \tanh({\pi \nu}  )  \ov  \pi\nu   } Q(\nu)  \, ,
\ee
where the $Q$-function is defined via
\be
  \nonumber
  Q(\nu) := \nu \cosh(\pi \nu) \psi(\nu) \, ,
\ee
if we apply the quantization condition $Q(0) = Q(\pm i) = 0$ along with certain analyticity conditions \cite{Fateev:2009jf}. We refer to finite difference equations like  \eqref{eq:fdifeq} as {\it quantum spectral curves} (QSCs). QSCs have been developed and effectively used to compute the spectrum \cite{Gromov:2013pga, Gromov:2014caa} in the AdS/CFT correspondence \cite{Maldacena:1997re, Gubser:1998bc, Witten:1998qj} in the planar limit. For recent reviews, see \cite{Gromov:2017blm, Kazakov:2018ugh, Levkovich-Maslyuk:2019awk}. Unlike QSCs in AdS/CFT, which describe infinitely many degrees of freedom, equation \eqref{eq:fdifeq} requires only $\mu$ to be determined by quantization conditions due to the rigidity of the two-fold string.

The goal of this paper is to find an alternative method for quantizing the yo-yo (or folded) string.
The issue with equation \eqref{eq:fdifeq} is that its classical limit is non-analytic, which is not natural from an integrability perspective. Indeed, for large $\nu$ we get
\be
  \nonumber
    { \pi\nu  \coth({\pi \nu}  )    } \ \to \  {\pi |\nu|} \, .
\ee
This non-analyticity directly arises from the absolute value in the potential term in equation \eqref{eq:hketto}, and it likely contributes to the difficulty in calculating the spectrum.
In AdS/CFT,  classical spectral curves are analytic, prompting the question of whether a similar analytic description is possible for the folded string, potentially yielding a simpler spectrum. As we will see, the answer is affirmative, though it comes with a price.

\clearpage

\section{The folded string in AdS$_2$}
\label{sec:folded}

In this paper, we choose the target space to be two-dimensional anti-de Sitter (AdS) space. Although the background curvature complicates the description, the AdS target space allows for a different approach to quantizing the string. Instead of considering two particles connected by a string as in \eqref{eq:hketto}, we primarily focus on a closed folded string, which has the same dynamics, with the folds acting as massless particles. For a depiction of the system, see Figure \ref{fig:yoyo}.
 The folded string is placed inside AdS$_2$, whose metric can be written in global coordinates as
\be
  \nonumber
    ds^2 = R^2 (-\cosh^2 \rho \, d\tau^2 + d\rho^2 ) \, , 
\ee
where $R$ is the AdS radius.
The action can be written as \cite{Ficnar:2013wba}
\be
  \nonumber
  \mathcal{S}_\textrm{string} = -{1 \over 4\pi\alpha'} \int_\text{w.s.} d^2 \sigma \, \sqrt{-h} h^{ab}
    \partial_a X^\mu \partial_b X^\nu G_{\mu\nu}  +
   \int_\text{folds} d\xi \, {1 \over 2\eta} \dot{X}^\mu \dot{X}^\nu G_{\mu\nu} \,,
\ee
where the first integral is over the worldsheet, and the second is over the locations of the two folds (or the boundary of half of the worldsheet). $\alpha'$ is the square of the string length, $X$ is the embedding function, $h$ is the worldsheet metric, $G$ is the target space metric, and finally $\eta$ is a Lagrange multiplier ensuring that the folds move with the speed of light.

Let us place the string centered at $\rho=0$ and parametrize its endpoints using the coordinate $\xi=\tau$.
The location of the endpoints is $\pm \tilde \rho(\tau)$. After eliminating $\eta$, the Hamiltonian is expressed  as \cite{Callebaut:2015fsa}
\be
  \label{eq:hami}
 \tilde H(\tilde p, \tilde \rho) = |\tilde{p}| \cosh \tilde \rho + {4g} \sinh |\tilde \rho |\, ,
\ee
where $\tilde{p}$ is the conjugate momentum, and the constant
\be
  \nonumber
  g:={R^2 \ov 2 \pi \alpha'}
\ee
has been introduced for convenience. In Sections \ref{sec:folded} and \ref{sec:tachyon} we will assume $g>0$ unless otherwise mentioned.

Hamilton's equations are
\bea
  \nonumber
  \dot{\tilde p} &=&      -|\tilde{p}| \sinh \tilde \rho - {4g} \cosh \tilde \rho \ \textrm{sgn} \, \tilde\rho  \, , \\
  \nonumber
  \dot{\tilde\rho}  &=&  \cosh \tilde \rho \ \textrm{sgn} \, \tilde p \, .
\eea
If the two folds collide at $t=0$, then we get the classical solution
\be
  \label{eq:classol}
  \tilde\rho(t) = \begin{cases}
  \text{arsinh}  \tan t \quad & \textrm{for} \ -{T \ov 4} \leq t<{T \ov 4} \, , \\
  \text{arsinh} \tan \le({T \ov 2} - t\ri)  \quad & \textrm{for} \ \ \ \ {T \ov 4} \leq t < {3T \ov 4}  \, ,
\end{cases}
\ee
\be
  \nonumber
  \tilde p(t) = \begin{cases}
  M \cos t - 4g \sin t \quad & \textrm{for} \ 0 \leq t<{T \ov 2} \, , \\
  M \cos\le({T } - t\ri) - 4g \sin \le({T } - t\ri)  \quad & \textrm{for} \ {T \ov 2} \leq t < {T}  \, ,
\end{cases}
\ee
where $M$ is the energy of the solution, and both functions are periodic $\tilde\rho(t) = \tilde\rho(t+T)$, $\tilde p(t) = \tilde p(t+T)$ with period
\be
  \nonumber
  T = 4\arctan {M \ov 4g}   \, , \qquad 0 < T < 2\pi \, ,
\ee
$\tilde\rho(t) $ is plotted in Figure \ref{fig:motion} (left).

\subsection{Classical spectral curve and its flat space limit}

The classical string motion in AdS is integrable and thus a spectral curve for the closed folded string can be computed with the result \cite{Vegh:2021jhl, Vegh:2021jqo, Vegh:2023snc}
\be
  \label{eq:csc}
   e^p + e^{-p} + 2 - { M^2 \ov u^2 -4g^2}  = 0 \, .
\ee
Here $M$ is the mass of the string, and $u$ and $p$ are canonical variables \cite{Dorey:2006mx, Vicedo:2008ryn},
\be
  \nonumber
  \{ u, \, p \} = 1 \, .
\ee
They are related to  $\tilde p$ and $\tilde\rho$ via  the somewhat complicated map
\bea
  \nonumber
  \tilde p &=& 2 \sinh\le( {p \ov 2} \ri) \sqrt{u^2 - 4g^2}\, , \\
  \label{eq:canonical}
  \tilde\rho  &=&  2 \,  \text{artanh} \le(e^{-{|p|\ov 2}} \sqrt{|u|-2g \ov |u|+2g}   \ri) \, \textrm{sgn} \, u \, .
\eea
\begin{figure}[h]
\begin{center}
\includegraphics[width=7.2cm]{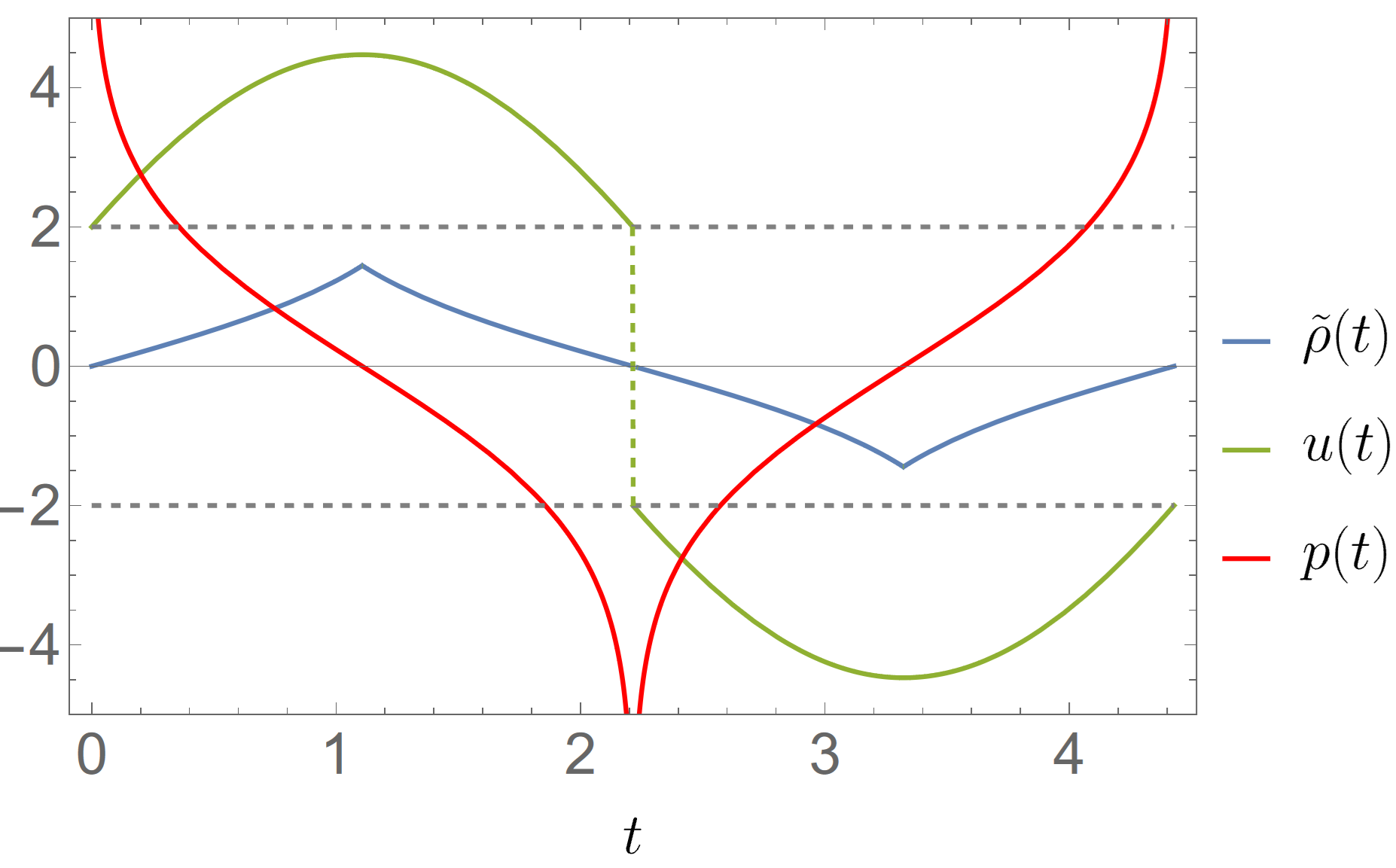} \qquad
\includegraphics[width=8cm]{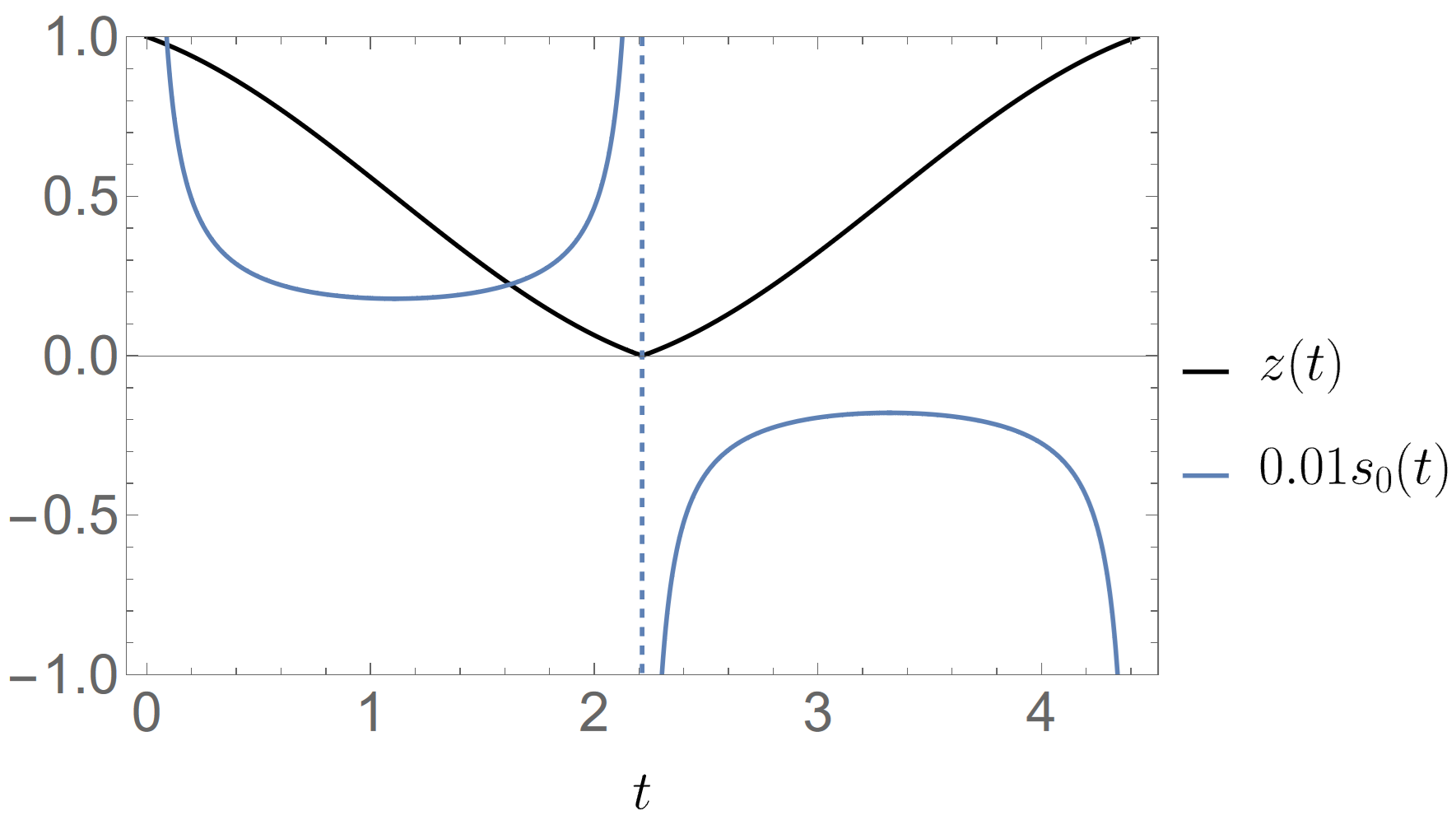}
\caption{\label{fig:motion}
Oscillatory motion of the folded string (for $g=1$, $M=8$). Endpoints move with the speed of light and they suddenly change direction. The momentum $p$ diverges when the folds (particles) collide at $\tilde \rho = 0$, and $u$ jumps from $2g$ to $-2g$ or vice versa. The momentum fraction variable $z$ oscillates between 0 and 1. The conjugate variable is $s_0 \sim \mathcal{O}(g)$. }
\end{center}
\end{figure}
This map is a canonical transformation away from the collision points, where $\tilde \rho = 0$ and $u= \pm 2g$. The physical configuration space is the $|u| \ge 2g$ region on the real $u$ line. Points on the interval $-2g < u < 2g$ do not correspond to points on the original phase space manifold, which was parametrized by $(\tilde p, \tilde \rho)$.
Interestingly, the map can be derived  by computing level sets, similar to how the Ruijsenaars-Schneider model \cite{RUIJSENAARS1986370} determines the motion of solitons in various (1+1)-dimensional integrable field theories. In this case, soliton positions are analogous to the physical coordinate $\tilde\rho$ of the fold \cite{Vegh:2023snc}.

\begin{figure}[h]
\begin{center}
\includegraphics[width=7cm]{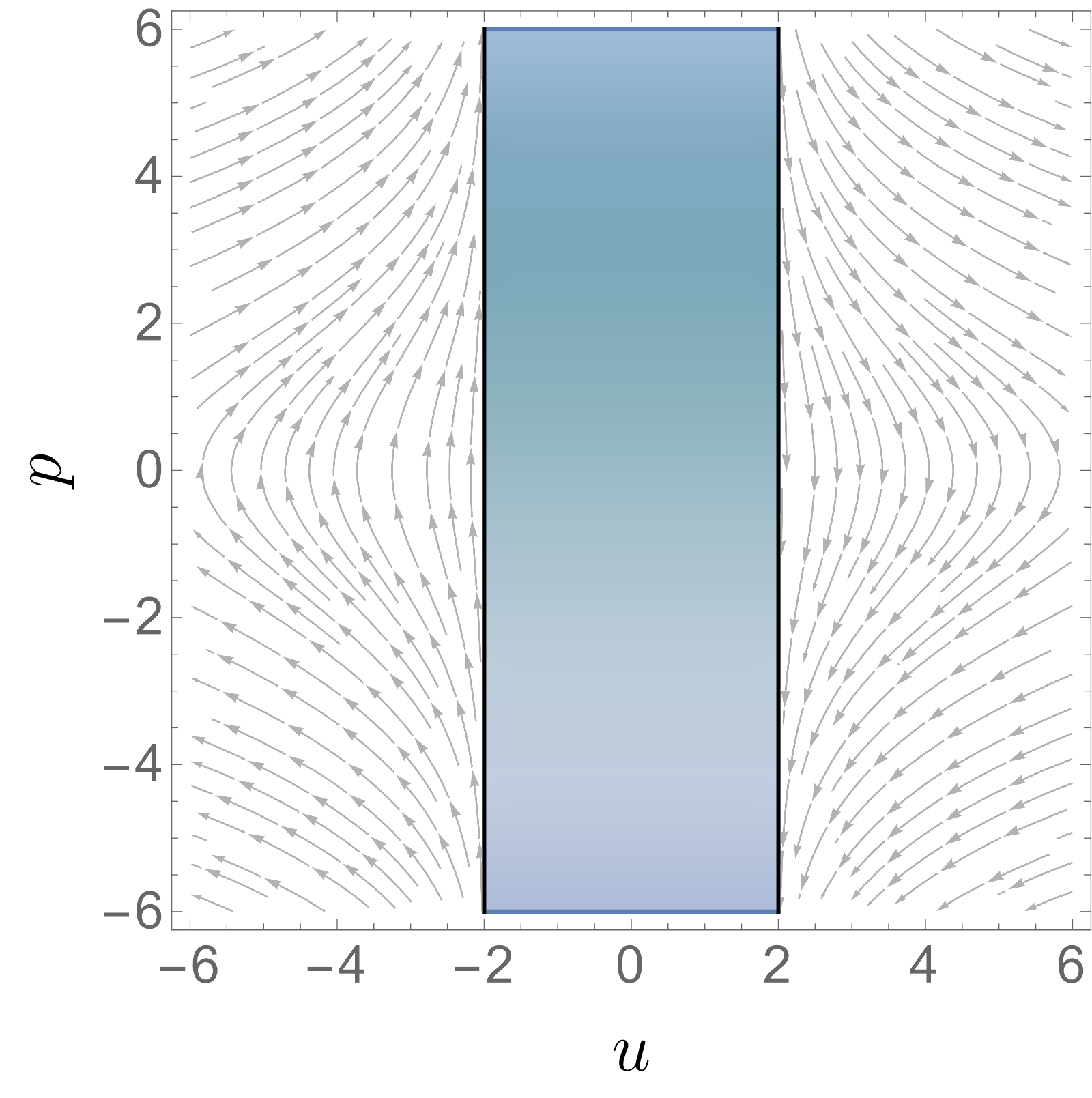} \qquad
\includegraphics[width=7cm]{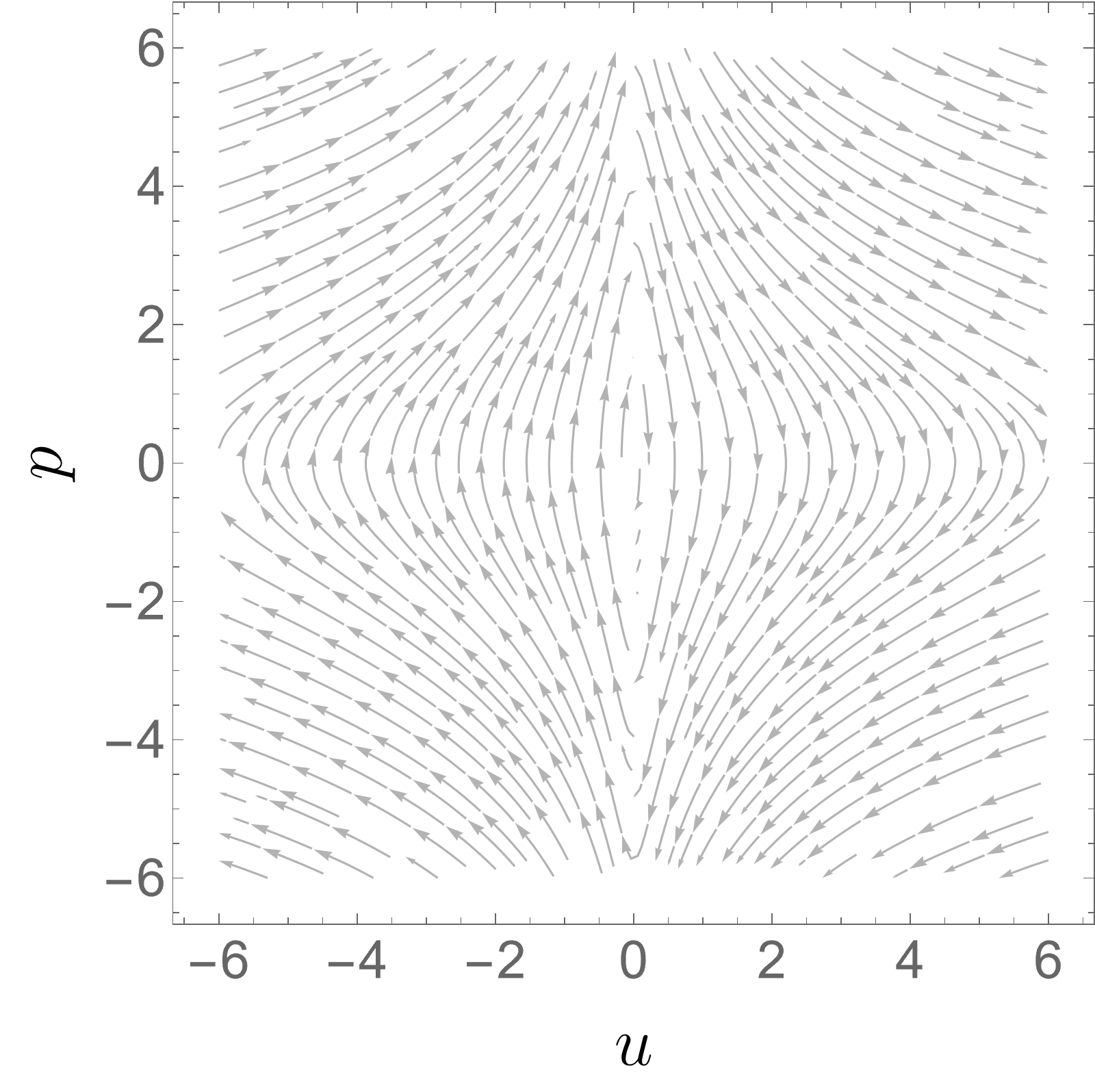}
\caption{\label{fig:stream}
{\it Left:} Phase space and Hamiltonian flow for $g=1$. The region $-2g<u<2g$ is excluded from the phase space. As $|u| \to 2g$ from above, $p$ runs out to infinity. {\it Right:} Same plot for $g={i\ov 8}$ (i.e. $\psyk=4$).
In this case, both $u$ and $p$ stay finite during the oscillation.
}
\end{center}
\end{figure}

In terms of the coordinates on the spectral curve we have
\be
  \nonumber
  M(u,p) =  \tilde H(\tilde p, \tilde \rho) \, ,
\ee
where $M$ on the left-hand side is expressed from \eqref{eq:csc},
\be
  \nonumber
   M(u,p) = 2 \cosh\le({p \ov 2}\ri) \sqrt{u^2 - 4g^2}   \, .
\ee
The motion can be computed from Hamilton's equation using the Hamiltonian $M(u,p)$. The corresponding Hamiltonian flow is plotted in Figure \ref{fig:stream} (left).
We get the solution
\bea
  \nonumber
  u(t) &=&  \begin{cases}
  2g\cos t+{M\ov 2} \sin t \qquad & \text{for} \ 0<t<{T \ov 2} \, ,\\
  -2g\cos(t-{T \ov 2})-{M\ov 2} \sin(t-{T \ov 2}) \qquad & \text{for} \ {T \ov 2}<t<T \, .
\end{cases}  \\
  \nonumber
  p(t) &=& 2 \, \text{arsinh} {u'(t) \ov \sqrt{u(t)^2 - 4g^2}}  \, .
\eea
The motion is displayed in Figure \ref{fig:motion} (left). At $t={T \ov 2}$, the two folds (particles) collide. In order to be able to map $(u(t), p(t))$ to the physical solution \eqref{eq:classol} via the map \eqref{eq:canonical}, we have to remove by hand the portions where $u(t) \in (-2g,2g)$ and glue the remaining pieces. If we perform this surgery, then $u(t)$ jumps from $+2g$ to $-2g$ (or vice versa at a later time) and $p(t)$ diverges as seen in the Figure.

To connect the spectral curve \eqref{eq:csc} to that of the \thooft equation, let us rescale the mass $\mu^2:={\pi M^2 \ov 4g}$.
This is the same $\mu$ that appears in the introduction, with  $\kappa=2g$.
Taking the $g\to \infty$ flat space limit while keeping $\mu$ fixed reduces equation \eqref{eq:csc} to
\be
  \label{eq:nonanal}
  e^p + e^{-p}  + 2 - {\mu^2 \ov \pi |\nu|}  = 0 \, ,
\ee
where we have introduced  a new coordinate
\be
  \label{eq:cutting}
  \nu := \begin{cases}
  u-2g \qquad & \textrm{for} \ u\geq +2g \, , \\
  u+2g \qquad & \textrm{for} \ u\leq -2g \, .
\end{cases}
\ee
This effectively removes the region $-2g < u < 2g$  and glues the two remaining half-lines.
The QSC in \eqref{eq:fdifeq}, which corresponds to the \thooft equation, is essentially the quantized version of \eqref{eq:nonanal}, with extra signs due to the $i$-antiperiodic factor in the definition of the $Q$-function.

\clearpage

\subsection{Momentum fraction variable}

The AdS analog of the momentum fraction variable $z$ in \eqref{eq:momfrac} can be obtained from $p$ via the same expression as in \eqref{eq:pdef},
\be
  \label{eq:canz}
  p = \log{z \ov 1-z} \,   .
\ee
From the integrability perspective, there are two natural choices for the canonical conjugate $s$, which we denote by $s_\pm$. They are related to $u$ via \cite{Vegh:2023snc}
\be
  \label{eq:cans}
  u =  \pm 2g+ s_\pm z(1-z) \, .
\ee
Physically allowed values are
 \be
  \label{eq:domainz}
   0 \leq z \leq 1  \,
\ee
and from $|u| \ge 2g$ we get
\be
  \label{eq:domains}
  s_+ \leq -{4g  \ov z(1-z)}
  \, , \qquad \ s_+ \geq 0 \, ,
\ee
or for the other choice
\be
  \label{eq:domains2}
  s_- \leq 0
   \, , \qquad \ s_- \geq  {4g  \ov z(1-z)} \, .
\ee
The mass squared of the string can be expressed in terms of $(z, s_+)$ or  $(z, s_-)$,
\be
 \label{eq:hamsz}
  M^2 =  \pm 4gs_\pm + z(1-z)s_\pm^2 \, .
\ee
Note that we cannot neglect the $z(1-z)s_\pm^2$ term when taking the $g \to \infty$ flat space limit in \eqref{eq:hamsz}. Although $s_\pm \sim \mathcal{O}(1)$ in a half-period, after the particles (folds) collide, in the other half-period we have $s_\pm \sim \mathcal{O}(g)$, which means that both terms on the RHS are $\mathcal{O}(g^2)$.

To simplify the flat space limit, we define the variable $s$ as
\be
  \label{eq:svar}
  s := \begin{cases}
  s_+ \quad & \textrm{for} \ s\geq 0 \, , \\
  s_- \quad & \textrm{for} \ s<0 \, .
\end{cases}
\ee
In terms of $s$, \eqref{eq:hamsz} becomes
\be
  \label{eq:adsham}
  M^2 =  4g|s| + z(1-z)s^2 \, .
\ee
The advantage of this form is that the $g \to \infty$ flat space limit is straightforward since the second term on the RHS is always subleading. After rescaling $M$, one arrives at \eqref{eq:infmom} with $\kappa=2g$. Thus, the \thooft equation ``solves'' the issue of taking the flat space limit by switching between the coordinates $(s_+, z)$ and $(s_-, z)$ at every collision of the particles or folds (when $s_\pm$ switches sign and $z=0$~or~$z=1$). This led to the absolute value in the potential in \eqref{eq:adsham}. On the $u$-plane switching between $s_\pm$ corresponds to the surgery in \eqref{eq:cutting}, which led to the absolute value in the spectral curve \eqref{eq:nonanal}.

\subsection{Quantization and the \teller potential}

Equation \eqref{eq:adsham} can be promoted to a \schr equation, which generalizes the 't Hooft equation to AdS space, and interpolates between evenly spaced conformal dimensions at $g=0$ and 't Hooft's nearly linear Regge trajectory in the $g \to \infty$ flat space limit (see \cite{Vegh:2023snc} for numerical results).

In the following, we will consider an alternative way to quantizing the system.
We will use $(s_\pm, z)$ and the analytic mass-squared expression in equation \eqref{eq:hamsz} as the starting point. For definiteness, let us consider the variable with the positive sign. The oscillation in the $s_+$ coordinate is asymmetric: a coherent state, represented  by the wavefunction $\varphi(t, z)$, might initially form a smooth peak, but upon reaching $z=0$ or $z=1$, it is reflected back as a highly oscillatory wavefunction.
To symmetrize the situation, let us introduce the variable $s_0$ defined by the shift
\be
  \nonumber
  s_+ = s_0 - {2g  \ov z(1-z)} \, .
\ee
In terms of $s_0$, equation \eqref{eq:hamsz} becomes
\be
  \label{eq:result}
  M^2 =  -4g^2\le({1\ov z}+{1\ov 1-z}\ri) + s_0 z(1-z)s_0 \, ,
\ee
where the (soon-to-be) operators in the last term on the right-hand side have been ordered in a specific way. (An alternative operator ordering will be discussed in the Discussion section.)
The corresponding \schr equation can be obtained by setting  $s_0 \to i \p_z$, and letting the terms act on a wavefunction $\varphi(z)$,
\be
  \label{eq:main}
  M^2 \varphi(z) =  -{4g^2 \ov z(1-z)}\varphi(z) + (2z-1)\varphi'(z) - z(1-z)\varphi''(z) \, .
\ee
We note that the non-trivial domain of $s_0$ has not been properly handled here. Classically, we must have
\be
  \label{eq:forbid}
  | s_0| \geq {2g  \ov z(1-z)} \, .
\ee
This also means that in the $g \to \infty$ flat space limit, the wavefunction $\varphi(z)$ is always highly oscillatory.
For now, let us proceed and revisit the issue of domains later.
Switching to the $p$ coordinate via \eqref{eq:canz}, we obtain the hyperbolic \teller equation,
\be
  \label{eq:pt}
  -\varphi''(p) - \frac{M^2}{4 \cosh^{2}\left(\frac{p}{2} \right)}\varphi(p) = 4g^2 \varphi(p) \,.
\ee
A solution is given by
\be
  \label{eq:ptsol}
  \varphi_A(p) = e^{2i g p} \, {}_2F_1\left(1 - h, h; \ 1 + 4i g; \ \frac{1}{1 + e^{-p}}\right) \, ,
\ee
where  the mass has been expressed in terms of the conformal dimension $h$ via
\be
  \nonumber
  M^2 = h(h-1) \, .
\ee
The other independent solution $\varphi_B$ is given by \eqref{eq:ptsol} with $g \to -g$.

Another form of the \schr equation is obtained by setting
\be
  \nonumber
    z = {1+ \sin \theta \ov 2} \, .
\ee
Upon redefining the wavefunction
\be
  \nonumber
  \varphi(\theta) = {1 \ov \sqrt{\cos \theta}} \tilde \varphi(\theta)
\ee
we obtain
\be
  \label{eq:pt2eq}
   \le(M^2 + {1\ov 4}\ri) \tilde\varphi(\theta)  = -\tilde \varphi''(\theta) - \frac{{1\ov 4} + 16g^2}{ \cos^{2}(\theta)}\tilde\varphi(\theta) \,.
\ee
For $g>0$ this is a strongly attractive \teller potential.

\subsection{Boundary conditions}
\label{sec:bc}

To compute the spectrum of the \schr equation in \eqref{eq:pt}, we need to impose appropriate boundary conditions at $z = 0$ and $z=1$, which correspond to $p = \pm \infty$. At these points the two particles (folds) collide, and the potential in equation \eqref{eq:pt2eq} becomes singular
\be
  V \approx {c \ov (\theta \pm \pi/2)^{2}} \, ,
\ee
where $c < -{1\ov 4}$ for $g>0$. This is below the critical value for inverse-squared potentials. In order to have a self-adjoint extension of the potential, one must set appropriate boundary conditions. For a detailed discussion of the special  $g=0$ case, see \cite{Sword:2024gvv}. Discussions on regularizing and renormalizing inverse-squared potentials can be found in \cite{case1950, Beane:2000wh, Bawin:2003dm, Hammer:2005sa, Gitman_2010, CAMARADASILVA2024169549}.

From the \schr equation, we can compute a conserved probability current $J$ given by
\be
  \nonumber
  J(z) = -i z(1-z) \le[ \varphi(z)^* \varphi'(z) - \varphi(z) \varphi'(z)^* \ri] \, .
\ee
In $p$-space, this current is expressed as
\be
  \nonumber
  J(p) = -i \le[ \varphi(p)^* \varphi'(p) - \varphi(p) \varphi'(p)^* \ri] \, .
\ee
As $p\to -\infty$, the wavefunction has the expansion
\be
  \nonumber
  \varphi(p) = c_A \varphi_A(p) + c_B \varphi_B(p) \stackrel{p\to -\infty}{\approx}  c_A e^{2i g p} + c_B e^{-2i g p} \, .
\ee
For unitary time evolution, the probability must not leak out at the boundaries, which requires
\be
  \nonumber
  \lim_{p\to -\infty} J(p)=0 \, .
\ee
This gives the condition
\be
  \nonumber
  c_A =  \lambda_- { c_B} \, ,
\ee
where
\bea
  \nonumber
 & |\lambda_-|  = 1     \quad & \textrm{for} \ g>0 \, , \\
 \label{eq:conditions}
  &   \lambda_- \in \RR \quad & \textrm{for} \ g \in i \RR \, , \ -\tfrac{1}{4}<|g|<\tfrac{1}{4} \, .
\eea
Fixing $\lambda_-$ gives a family of boundary conditions.
Note that both independent solutions are normalizable when $-\tfrac{1}{4}<\Im g <\tfrac{1}{4}$. At  $g = \pm {i\ov 4}$, one of the two solutions becomes non-normalizable, making it impossible to impose the general boundary condition as described above.

Similarly, as $p \to \infty$, the wavefunction has the expansion
\be
  \nonumber
  \varphi(p)   \stackrel{p\to \infty}{\approx}  \tilde c_A e^{2i g p} + \tilde c_B e^{-2i g p} \, ,
\ee
where the coefficients are related to those at $p \to -\infty$ via
\be
  \label{eq:scatter}
  \binom{\tilde c_A}{\tilde c_B} =
{1 \ov \sinh(4 \pi  g)}
\left(
\begin{array}{cc}
 \frac{4 \pi  g   \Gamma (4 i g)}{\Gamma (1-4 i g) \Gamma (1-h+4ig) \Gamma (h+4ig)} & -i  \sin (\pi  h) \\
 i   \sin (\pi  h) & \frac{4 \pi  g  \Gamma (-4 i g)}{\Gamma (1+4 i g) \Gamma (1-h-4 i g) \Gamma (h-4 i g)} \\
\end{array}
\right)
\binom{c_A}{c_B} \, .
\ee
Requiring  $\lim_{p\to \infty} J(p)=0$ gives the condition
\be
  \nonumber
  \tilde c_A =  \lambda_+ {\tilde  c_B} \, ,
\ee
and as in \eqref{eq:conditions},
\bea
  \nonumber
 & |\lambda_+|  = 1     \quad & \textrm{for} \ g>0 \, , \\
  \nonumber
  &   \lambda_+ \in \RR \quad & \textrm{for} \ g \in i \RR \, , \ -\tfrac{1}{4}<|g|<\tfrac{1}{4} \, .
\eea
Fixing $\lambda_+$ introduces another parameter, resulting in a two-parameter family of boundary conditions parametrized by $\lambda_\pm$. The eigenfunctions of the \schr equation are orthogonal with respect to the measure
\be
  \nonumber
  d\mu = dz = {dp \ov 4 \cosh^2{( p/ 2)}} = \half \, d\theta \cos \theta \, .
\ee
The boundary conditions ensure that time evolution is unitary and that the eigenvalues of the \schr equation \eqref{eq:main} are real.

\begin{figure}[h]
\begin{center}
\includegraphics[width=7cm]{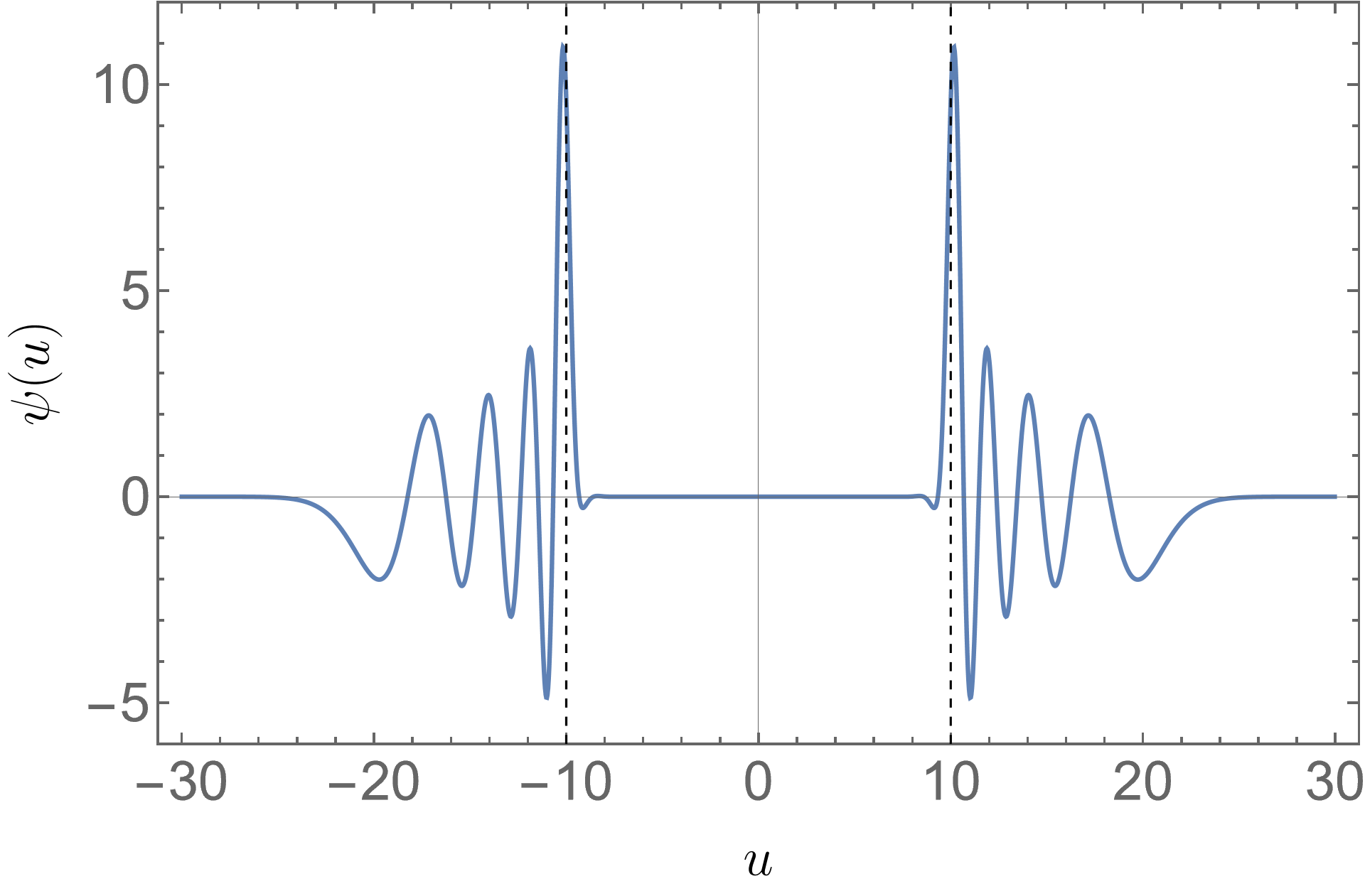} \qquad
\includegraphics[width=7cm]{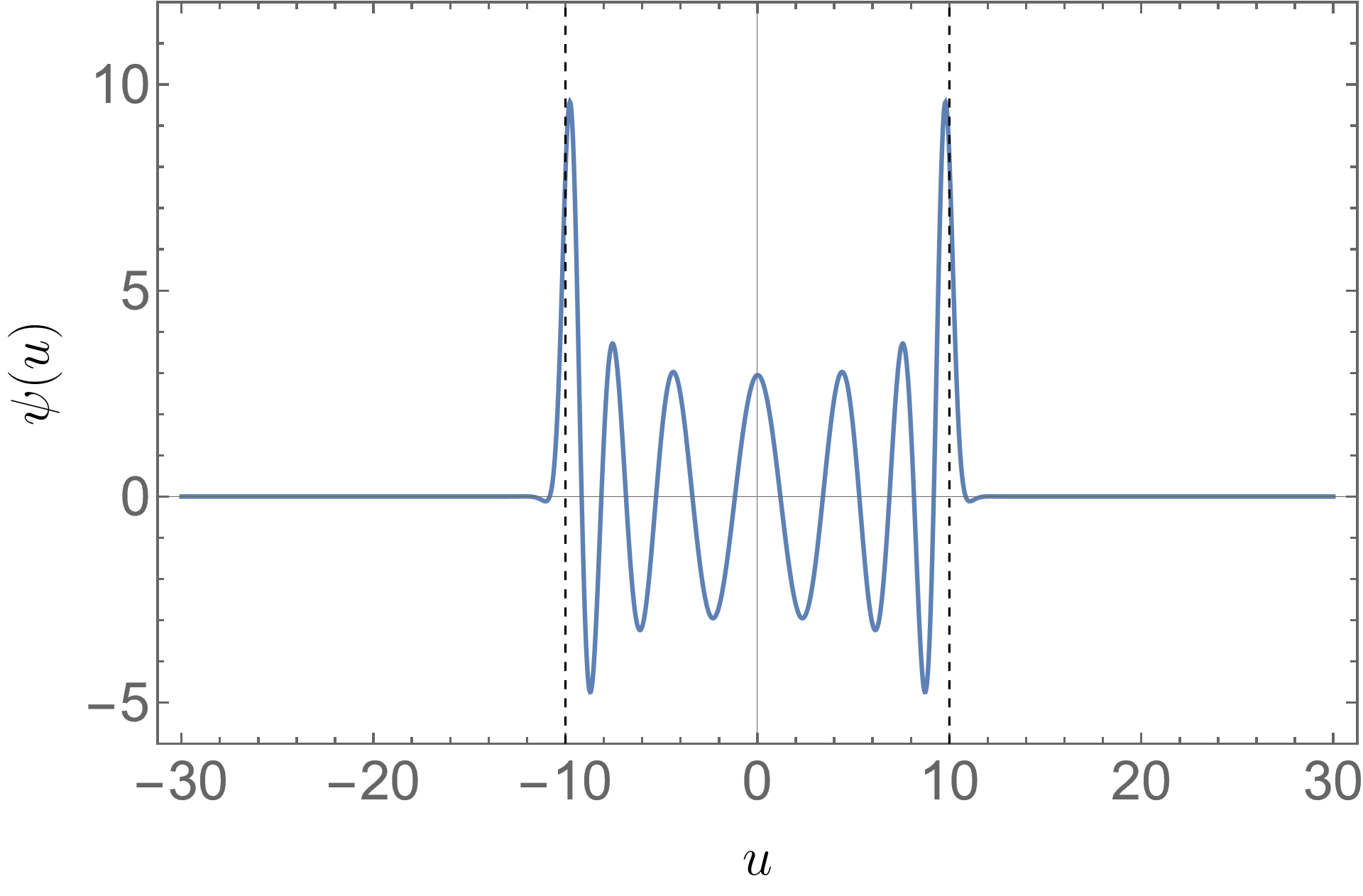}
\caption{\label{fig:wfu}
Example wavefunctions at $g=5$ with boundary conditions $\lambda_\pm = 1$. The first plot shows the eigenfunction at $h\approx 37.7047$. The second plot shows a tachyonic eigenstate at $h\approx \half + 24.548 i$. Tachyonic states are localized inside the forbidden region $-2g<u<2g$.
}
\end{center}
\end{figure}

\subsection{Quantum spectral curve}

The \schr equation \eqref{eq:main} can be transformed into a quantum spectral curve (a finite difference equation). The derivation is simpler than that of \eqref{eq:fdifeq} in \cite{Fateev:2009jf}. Defining
\be
  \nonumber
  D := 2 \cosh {p\ov 2} \, , \qquad \tilde D := \sqrt{u^2-4g^2} \, ,
\ee
the \teller equation \eqref{eq:pt} can be written as
\be
  \nonumber
  \tilde D^2 \varphi = M^2 D^{-2} \varphi
\ee
Multiplying the equation from the left by $D^2$ and defining $Q_0 := \tilde D^2 \varphi $, we get
\be
  \label{eq:qqq}
  D^2 Q_0 = M^2 \tilde D^{-2} Q_0
\ee
This can be written explicitly as
\be
  \label{eq:qsc0}
    Q(u+i)+Q(u-i) -2 Q(u)+  {M^2   \ov u^2 -4g^2  } Q(u) =0  \, ,
\ee
which is nothing but the quantum version of the classical spectral curve \eqref{eq:csc}.
Here we have included an $i$-periodic factor in the $Q$-function by setting $Q:= \sinh(\pi u) Q_0$, which multiplied the shifted terms $Q(u\pm i)$ by $-1$.

The wavefunctions \eqref{eq:ptsol} can be (inverse) Fourier transformed into $u$-space,
\be
\nonumber
  \psi_A(u) = \frac{
    \Gamma(2i g - i u) \Gamma(-2i g + i u) \Gamma(2 i g + i u) \,
}{
    \Gamma(h-1) \Gamma(1 + 2i g - h + i u)
}
{}_3F_2\left(\begin{matrix*}[r]
        1 - h, \,  1 + 4i g - h, \,    1 + 2i g + i u \\
         \hskip 0.3cm  1 + 4i g, \,    1 + 2i g - h + i u
    \end{matrix*} \ \Big| \, 1 \right)
\ee
and the other independent solution $\psi_B$ is given by the same expression with $g \to -g$. These functions have infinitely many poles, but no branch cuts. The wavefunction
\be
  \nonumber
  \psi(u) = c_A \psi_A(u)+ c_B \psi_B(u) \,
\ee
has poles at $u = \pm 2g -i$ and the coefficients can be read off from the residues
\be
  \label{eq:residues}
  \text{Res}(\psi, 2g-i) = c_A {h(h-1) \ov i-4g}   \, , \qquad
  \text{Res}(\psi, -2g-i) = c_B {h(h-1) \ov i+4g}   \, .
\ee
Finally, it can be checked that the $Q$-function
\be
  \nonumber
  Q(u) = (u^2 -4g^2)\sinh(\pi u) \psi(u) \,
\ee
satisfies the finite difference equation \eqref{eq:qsc0}.

\section{The spectrum for $g>0$}
 
\label{sec:tachyon}

Using the matrix in equation \eqref{eq:scatter}, the spectrum of the \teller equation can be calculated. We get
\bea
  \nonumber
  && \frac{4 \pi  e^{i \alpha_-} g \Gamma (4 i g)}{\Gamma (1-4 i g) \Gamma (1-h+4ig) \Gamma (h+4ig)}-\frac{4 \pi  e^{i \alpha_+} g \Gamma (-4 i g)}{\Gamma (1+4 i g) \Gamma (1-h-4 i g) \Gamma (h-4 i g)}= \\
  \label{eq:codi}
  && i \left(1+e^{i (\alpha_- +\alpha_+)}\right) \sin (\pi  h)  \, ,
\eea
which is an implicit equation for the eigenvalue $h$. Here, we emphasize that the boundary condition parameters are phases by expressing them as $\lambda_\pm = e^{i \alpha_\pm}$.
An example eigenfunction is shown in Figure \ref{fig:wfu} (left). In $u$-space, the forbidden region \eqref{eq:forbid} is simply the interval
\be
  \label{eq:forbidden}
  -2g < u < 2g \, ,
\ee
and the wavefunction in the Figure is localized mainly outside this region.
The corresponding classical motion $u(t)$ is plotted in Figure \ref{fig:motion} in red (for $g=1$).

It can be shown that for $g>0$, $M^2$ is not bounded from below: in addition to the $M>0$ states, the spectrum also contains infinitely many tachyons (see Figure \ref{fig:wfu}, right).
Their appearance is not surprising, since we have not explicitly excluded the forbidden region when we naively quantized the system.

\begin{figure}[h]
\begin{center}
\includegraphics[width=7.0cm]{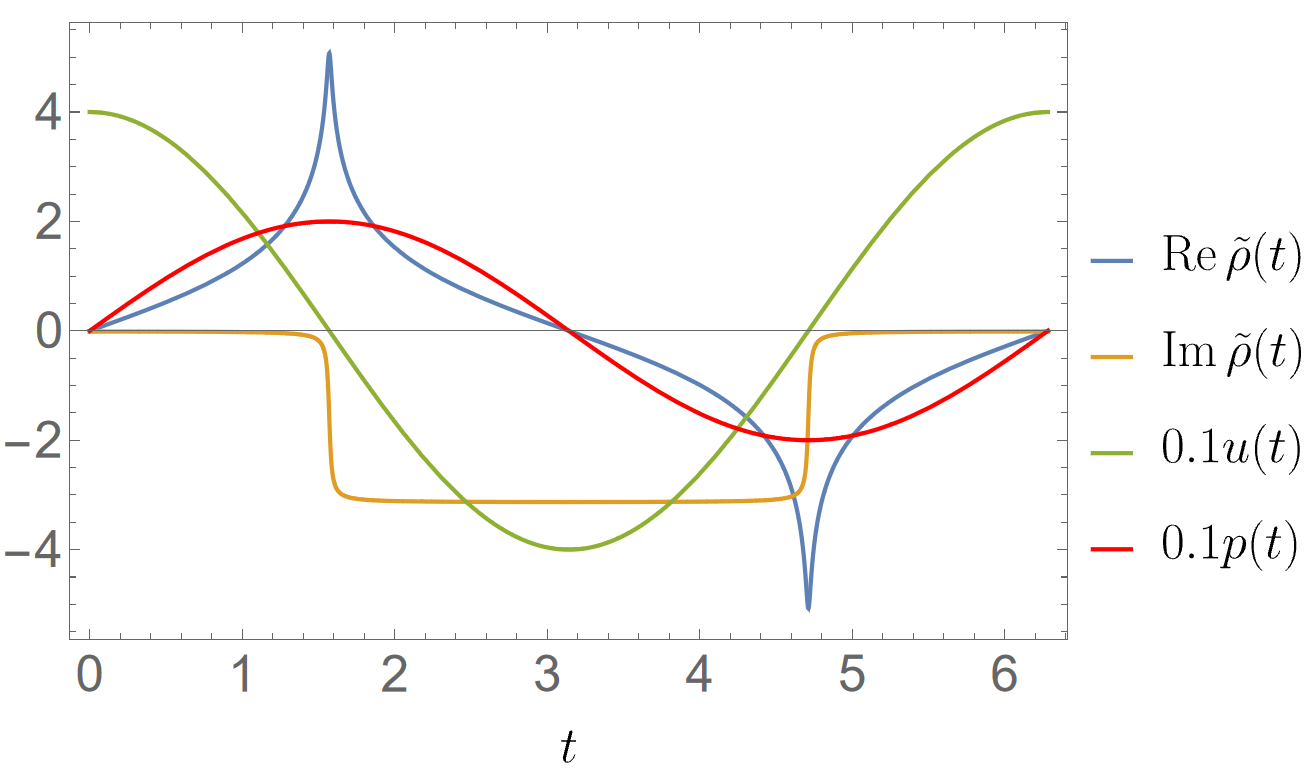} \qquad\qquad
\includegraphics[width=6.9cm]{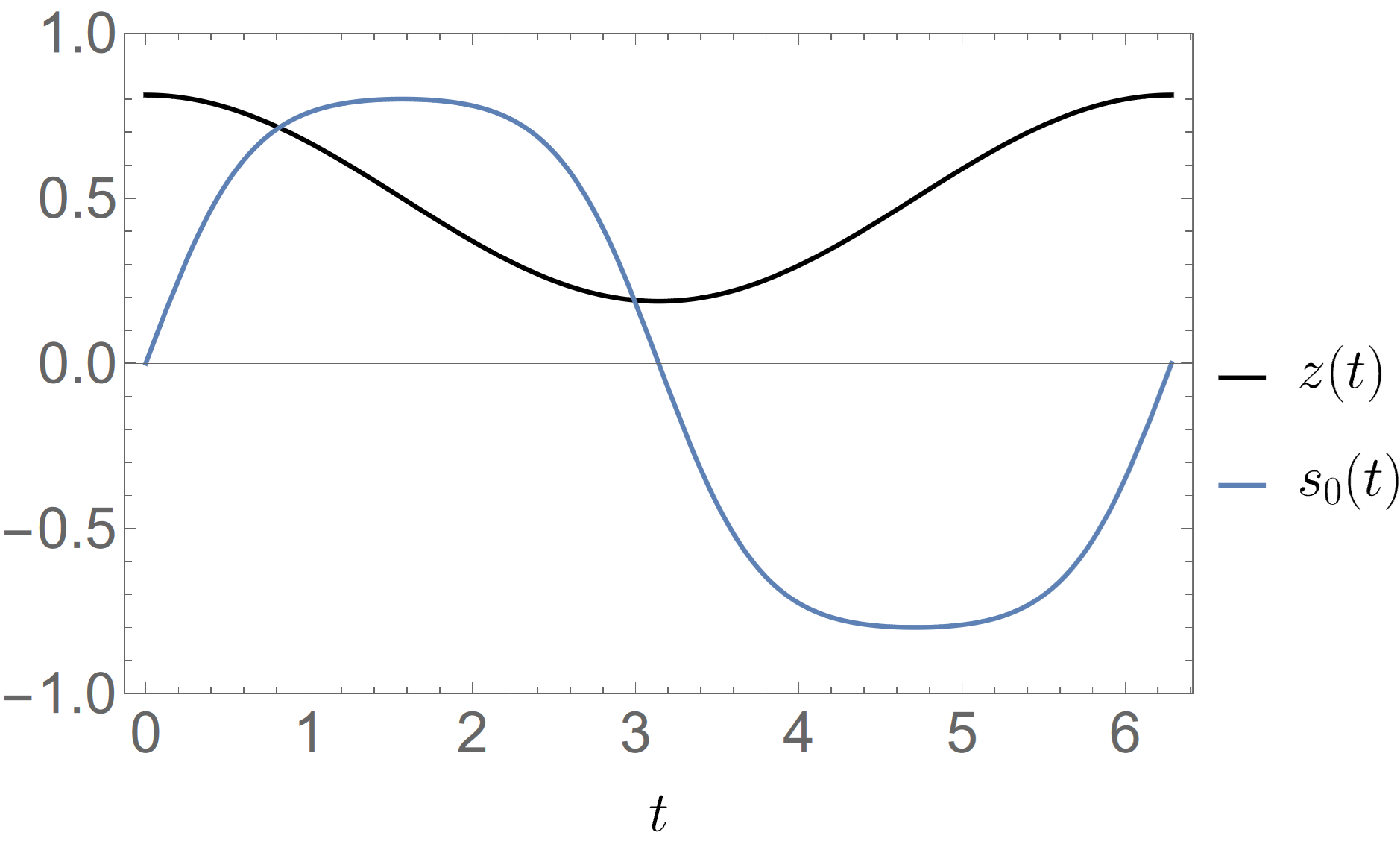}
\caption{\label{fig:motion2}
Classical motion for $g = {i \ov 8}$ (i.e. $\psyk=4$). Only $\tilde \rho(t)$ has a non-zero imaginary part. The diagrams are plotted with the amplitudes $u_0 = 20$ (left) and $u_0 = 0.2$ (right).
}
\end{center}
\end{figure}

Setting the boundary conditions $\alpha_\pm = 0$ and taking the $g \to \infty$ flat space limit, the condition \eqref{eq:codi} reduces to
\be
  \nonumber
  \sin {h(h-1) \ov 4g} = 0 \, ,
\ee
and thus the spectrum is an exactly linear Regge trajectory
\be
  \nonumber
   {M^2 } = h(h-1) =  4 \pi g  n \, , \qquad n \in \ZZ \, .
\ee
Although it is indeed simpler than the almost linear Regge trajectory in the spectrum of the massless \thooft equation, it contains tachyons. It may be possible to project out these states by carefully defining the `momentum' operator as discussed in \cite{Al-Hashimi:2021tkf}. However, such projections tend to result in complicated spectra, and we will not explore them further here.

\section{The spectrum for  $g \in (-{i\ov 4}, {i\ov 4}) $}
\label{sec:imag}
 
We will now explore what happens when $g$ takes on imaginary values. In this case, the forbidden region \eqref{eq:forbidden} disappears, and the configuration space becomes  the entire $u \in \RR$ line.
Although it is not entirely meaningful to talk about classical motion when the AdS radius is of $\mathcal{O}(\sqrt{\alpha'})$, we can still compute a classical solution
\bea
  \nonumber
  u(t) &=& u_0 \sin(t-t_0) \, , \qquad u_0, t_0 \in \RR \, , \\
  \nonumber
   p(t) &=& 2 \, \text{arsinh} {u'(t) \ov \sqrt{u(t)^2 - 4g^2}}  \, .
\eea
During the oscillation, both $u(t)$ and $p(t)$ remain real. Unlike in the case of $g>0$, the period of oscillation here is always $T=2\pi$.
Since $p(t)$ never diverges, the momentum fraction variable $z={1 \ov 1+ e^{-p}} $ oscillates within $[0,1]$, never reaching the boundaries (see Figure \ref{fig:motion2}, right). This implies that the two particles (or folds) never actually collide, eliminating the need for the absolute value in the second term of \eqref{eq:hami}.

The solution can be mapped to some kind of ``physical'' coordinates $(\tilde \rho, \tilde p)$ by choosing a single analytic branch of \eqref{eq:canonical}, e.g. by removing the absolute values in the expression. We get
\bea
  \nonumber
  \tilde\rho(t) &=& \log{ \sqrt{u_0^2 - 4g^2} + u_0 \sin(t-t_0)\ov 2g + u_0 \cos(t-t_0) }  \, , \\
  \nonumber
  \tilde p(t) &=&  2 u_0 \cos(t-t_0)   \, .
\eea
The variable $\tilde\rho(t)$ acquires an imaginary part (depicted by the yellow line in Figure \ref{fig:motion2}).
The  Hamiltonian in \eqref{eq:hami} is replaced by
\be
 \nonumber
 {\tilde H}'  = \tilde{p} \cosh \tilde \rho + {4g} \sinh \tilde \rho \, ,
\ee
which is conserved during the motion. The interpretation of the system in these coordinates is less clear, since $g$ is imaginary and $\tilde \rho(t) \in \CC$, and thus we will not explore it further. Instead, we will now discuss the eigenvalue spectrum of the \schr equation \eqref{eq:main}.

\subsection{The Sachdev-Ye-Kitaev spectrum}

We will consider $g$ in the imaginary range
\be
  \nonumber
  g = i g_0 \, , \qquad -{1\ov 4}< g_0 < {1\ov 4} \, ,
\ee
and parametrize it by
\be
  \nonumber
  g = {i \ov 4}(1-2 \Delta) \, , \qquad \Delta \equiv {1\ov \psyk} \, , \qquad 1 < \psyk < \infty \, .
\ee
Surprisingly, with appropriate boundary conditions and antisymmetrization, the resulting spectrum exactly matches that of the disorder-averaged Sachdev-Ye-Kitaev (SYK) model with $\psyk$-fermion interactions \cite{1993sachdev, kitaev, Polchinski:2016xgd, Maldacena:2016hyu}. The SYK theory is a simple model of holography, consisting  of $N$ Majorana fermions $\psi_i$. There is no spatial dimension, only time, and the fermions are coupled via random 4-fermion interactions. One considers an ensemble of theories whose members have Hamiltonians of the form
\be
  H = i^{\psyk/2} \sum_{1 \leq i_1 < i_2 < \ldots < i_\psyk \leq N}  J_{i_1\ldots i_\psyk} \psi_{i_1} \cdots \psi_{i_\psyk} \, ,
\ee
where $\psyk$ is even, and the couplings $J_{i_1\ldots i_\psyk} $ are drawn from a random Gaussian distribution. These couplings are fixed for each  member of the ensemble, and physical quantities are computed by averaging over the ensemble. The dimensions of the fermions are given by $\Delta$ in the infrared. For an introduction to SYK, see the reviews \cite{Sarosi:2017ykf, Rosenhaus:2018dtp}.

\begin{figure}[h]
\begin{center}
\includegraphics[width=8cm]{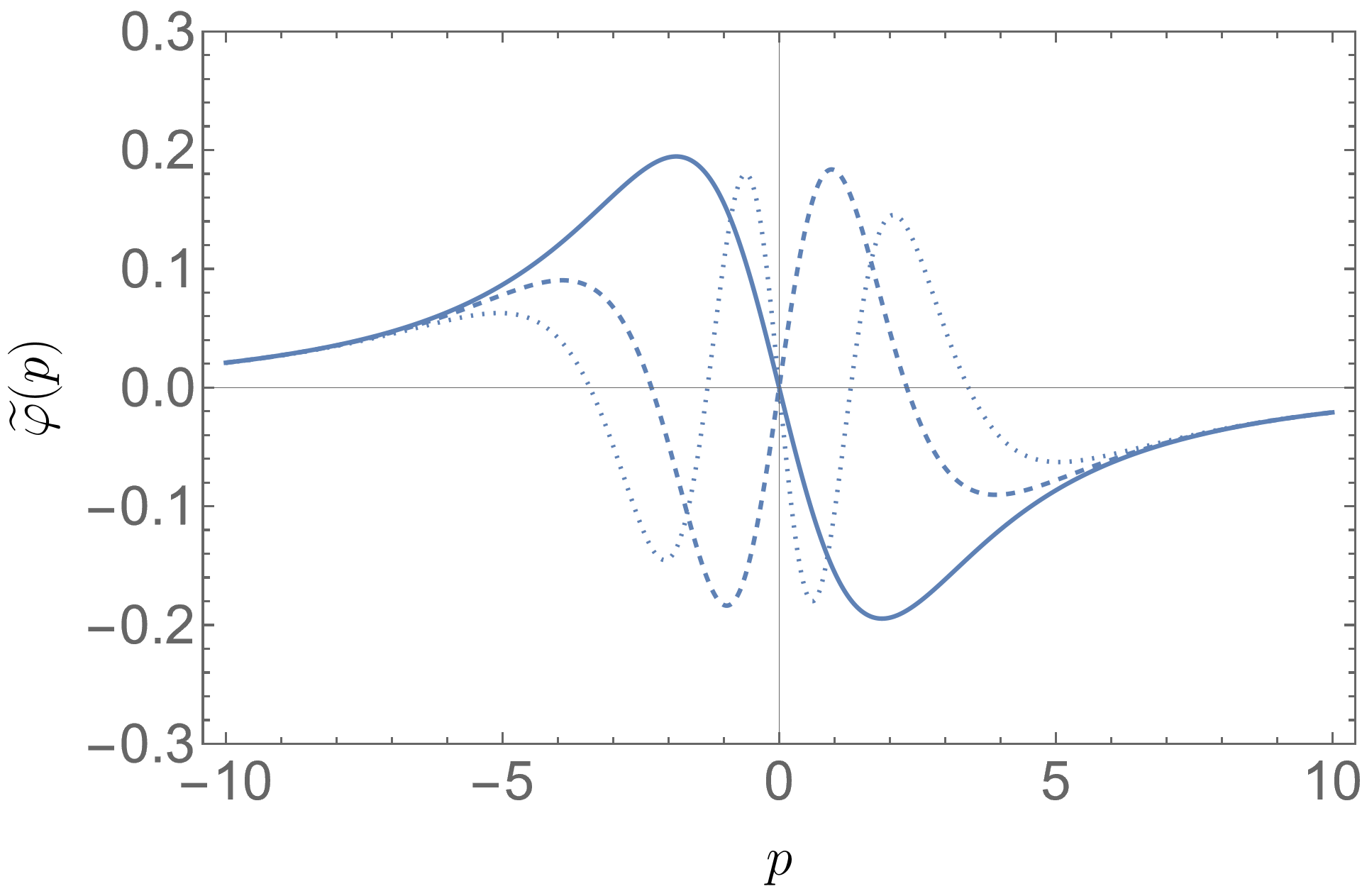}
\caption{\label{fig:wfp}
The first three $\Delta={1\over 4}$ ($g={i\over 8}$) wavefunctions. The functions have been rescaled: $\tilde\varphi(p) = {\varphi(p) \ov 2 \cosh({p\ov 2})}$ so that $|\tilde\varphi(p)|^2$ is the probability density. }
\end{center}
\end{figure}

  The SYK model is conjectured to have a gravity dual defined on AdS$_2$.
It has been proposed  (see Section 6 of \cite{Maldacena:2016hyu} or Section 5 of \cite{Rosenhaus:2018dtp}) that the bulk dual is some kind of string theory, possibly  related to the \thooft model, with low-tension strings $\ell_\text{string} \sim R_\text{AdS}$. (For other proposals, see \cite{Gross:2017hcz, Das:2017pif, Goel:2021wim}.) In the following, we show that the SYK spectrum of bilinear fermion operators can be obtained from the simple yo-yo string when the radius squared of the AdS target space is set to specific imaginary values.

In the  dual of SYK, two fermions are thought to be connected by a flux tube, and the state should be antisymmetric under the exchange of the two particles. This can be implemented in our system by antisymmetrizing the wavefunction $\varphi(p)$, which eliminates half of the eigenvalues. If $\varphi(p) $ is a solution to \eqref{eq:pt}, then so is $\varphi(-p)$. The antisymmetrized wavefunction
\be
  \nonumber
  \varphi_a(p) := \varphi(p) - \varphi(-p)
\ee
has the expansion
\be
  \nonumber
  \varphi_a(p)\approx  \begin{cases}
     \  c_A e^{(\Delta - \half)p} + c_B e^{(\half-\Delta)p} \quad  & \textrm{for} \ p\to -\infty \, , \\
  \ \tilde c_A e^{(\Delta - \half)p} +\tilde c_B e^{(\half-\Delta)p} \quad & \textrm{for} \ p\to +\infty \, .
\end{cases}
\ee
Due to the antisymmetrization, $\tilde c_A = -c_B$ and $\tilde c_B = -c_A$.
Equation \eqref{eq:scatter} also relates $c_i$ and $\tilde c_i$, from which we get
\be
  \label{eq:ab}
  {c_A\ov c_B} = {\Delta \ov 1-\Delta} k_\Delta(h)
\ee
where we have defined the function $k_\Delta$ \cite{Kitaev:2017awl}
\be
\nonumber
k_\Delta(h) :=  {f_\Delta(h) \ov f_\Delta(2)} \, , \qquad f_\Delta(h) := {g_\Delta(h) \ov g_{1-\Delta}(h)}
\, , \qquad g_\Delta(h) := {\Gamma(\Delta+  \tfrac{h}{2})\Gamma(\Delta+ \tfrac{1-h}{2})  } \, ,
\ee
so that it coincides with the eigenvalue of the conformal kernel that generates ladder diagrams in SYK \cite{Maldacena:2016hyu}.
There is a one-parameter family of boundary conditions defined by $\lambda_- $ and we have  $\lambda_+ = \lambda_-^{-1}$.
If we want the $h=2$ mode to be present in the spectrum, then we need to fix  \eqref{eq:ab} by plugging $h=2$ into the argument of $k_\Delta$. Then, since $k_\Delta(2) = 1$, the boundary condition
\be
  \label{eq:bdyfin}
  \lambda_- = {1 \ov \lambda_+} = {\Delta \ov 1-\Delta}
\ee
yields a string spectrum that matches the spectrum of SYK with $\psyk$-fermion interactions.
On the $u$-plane this boundary condition is equivalent to equating the residues of the wavefunction $\text{Res}(\psi, 2g-i)= \text{Res}(\psi, -2g-i)$. For a detailed analysis of the spectrum, refer to \cite{Maldacena:2016hyu}.

Finally we note that for the folded string configuration the wavefunction need not be antisymmetrized. In this case one obtains additional eigenvalues corresponding to symmetric wavefunctions, which can be calculated using \eqref{eq:codi}.

\subsection{Zero string tension ($\psyk = 2$)}

The case of $\psyk=2$ is a special case of the SYK model (see Appendix C in \cite{Gross:2016kjj} for a discussion), because it consists of free fermions with a random mass matrix, producing an evenly spaced spectrum.  This coincides with the string spectrum on a zero radius AdS space. Since we have  $k_\half(h) = -1$ identically, the spectrum has to be computed either by perturbing $\Delta$ away from $\half$ or by solving the \schr equation \eqref{eq:main} directly at $g=0$. We get the (normalized) wavefunctions
\be
  \nonumber
  \varphi_h(z) = \sqrt{2h-1} P_{h-1}(2z-1) \, , \qquad h = 1, 2, \ldots \, ,
\ee
where $P_k$ are Legendre polynomials. This solution is obtained by imposing the simplest boundary conditions
\be
  \nonumber
  \lim_{z \to 0,1} z(1-z)\varphi'(z) = 0 \, ,
\ee
which is discussed in Section 2 of \cite{Sword:2024gvv} in detail. Projecting on antisymmetric wavefunctions we finally get $h = 2, 4, 6, \ldots \, . $

\subsection{SYK with time-dependent couplings}
Boundary conditions other than \eqref{eq:bdyfin} produce the spectra of SYK models with time-dependent couplings. These models are discussed in Appendix H of \cite{Maldacena:2016hyu}. The authors promote the couplings of the SYK model  to time dependent fields with a two-point function given by
\be
  \nonumber
  \langle J_{i_1\ldots i_\psyk}(t) J_{i_1\ldots i_\psyk}(0) \rangle \propto {1 \ov |t|^{2 \alpha}} \, ,
\ee
where $\alpha \geq 0$ is a fixed parameter. In the $\alpha\to 0$ limit one recovers the original SYK model (to leading order in $N^{-1}$) with emergent conformal symmetry in the infrared. The eigenvalues of the kernel in the Schwinger-Dyson equation are computed to be \cite{Maldacena:2016hyu}
\be
  \nonumber
  \hat k =   {\psyk-1 \ov {1 \ov \hat \Delta } - 1} k_{\hat \Delta}(h) \, , \qquad \text{where} \ \hat \Delta := {1- \alpha \ov \psyk} \, ,
\ee
and the spectrum in this case is obtained by setting $\hat k  = 1$. We obtain the same spectrum from the yo-yo string by setting
\be
  \nonumber
   g = {i \ov 4}(1-2 \hat \Delta) \,   \qquad \text{and} \qquad  \lambda_- = {1 \ov \lambda_+} = {\Delta \ov 1-\Delta} \,  , \qquad \text{where} \ \hat \Delta := {1- \alpha \ov \psyk} \, ,
  \  \Delta := {1 \ov \psyk} \, ,
\ee
which is different from \eqref{eq:bdyfin}, because $g(\Delta)$ has been changed.

\section{Discussion}

In this paper, we quantized the single oscillatory degree of freedom of a closed folded string (or yo-yo string) in an AdS$_2$ target space of radius $R$. Quantization in terms of the physical variables $(\tilde p, \tilde \rho)$  in the center-of-mass  frame  fails due to  anomalies appearing in the target space symmetry algebra \cite{Lenz:1995tj}. Quantization using the classically equivalent pair $(s,z)$ yields  the (generalized) \thooft equation. Here, we employed the variables motivated  by integrability, namely $(u,p)$ (and the related $(s_0, z)$ pair), which also serve as coordinates on the spectral curve. In these variables, the Hamiltonian does not have an absolute value in the potential. However, this description is not fully equivalent to the original one, as the map between the variables ceases to be a canonical transformation when the folds (or particles) collide.
A complication arises for $g \equiv {R^2 \ov 2\pi \alpha'} >0$, where a forbidden region $-2g < u < 2g$ emerges. This must be excluded from the phase space, as it does not correspond to points on the physical phase space. As discussed in Section \ref{sec:tachyon}, naive quantization without addressing  this issue leads to tachyons localized in the forbidden region. It may be possible to project these states out by carefully redefining the operators, as suggested in \cite{Al-Hashimi:2021tkf}.

\begin{figure}[h]
\begin{center}
\includegraphics[width=4.5cm]{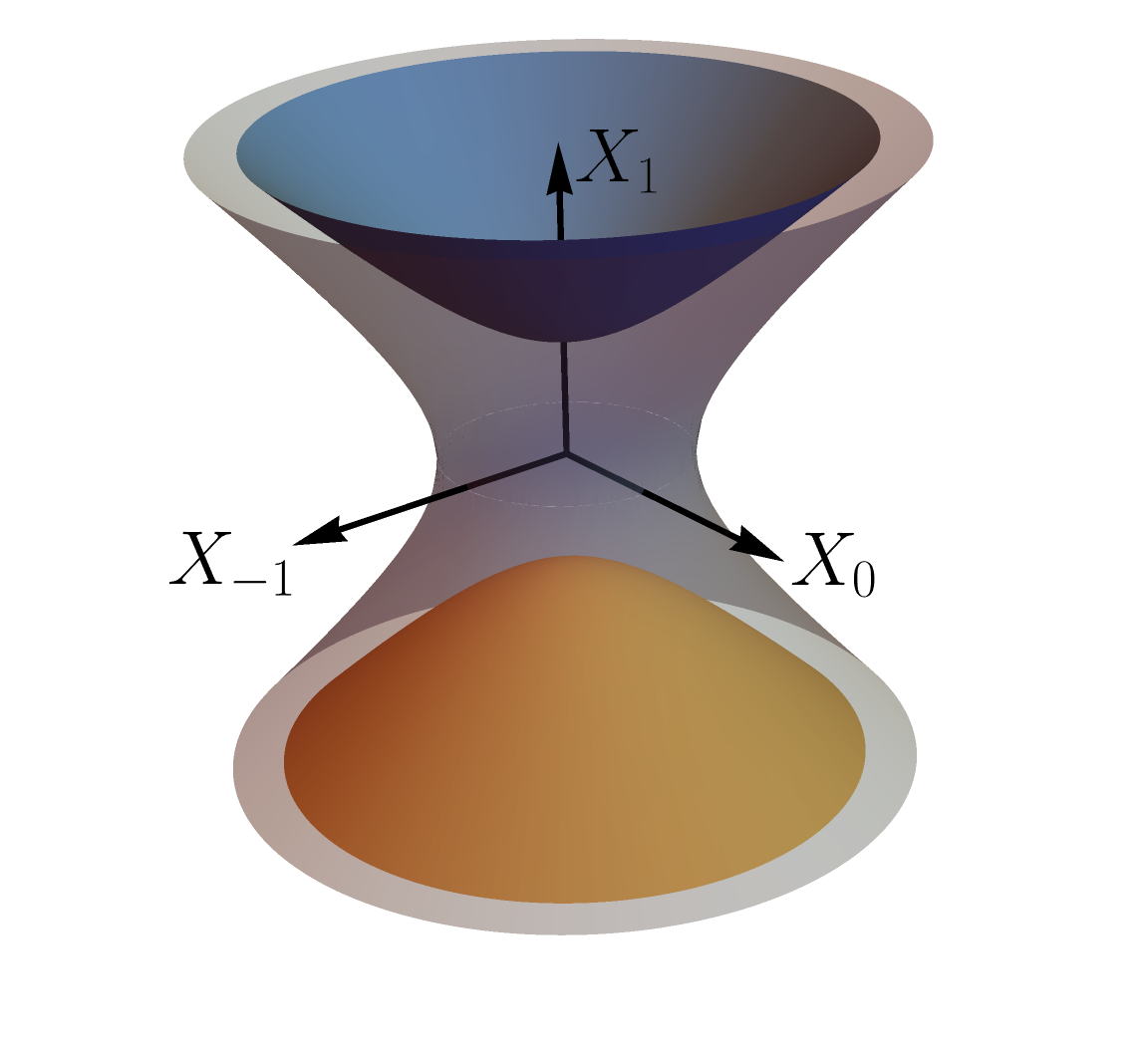}
\caption{\label{fig:hyper}
One- and two-sheeted hyperboloids in $\RR^{2,1}$ defined by $-X_{-1}^2-X_{0}^2+X_{1}^2 = \mp R^2$. The global time coordinate of AdS$_2$ (minus sign on the RHS in the equation) is the angle on the $X_{-1}, X_0$ plane, which has to be ``unwrapped.''
}
\end{center}
\end{figure}

We also considered the case   $g \in [0, {i\ov 4}) $, which proved more fruitful. With positive $\alpha'$, this case naively corresponds to a target space with an imaginary radius-squared, i.e. the Euclidean hyperbolic plane, illustrated   in Figure \ref{fig:hyper} as a two-sheeted hyperboloid (blue and orange sheets). However, this interpretation of the target space is not meaningful, as the ``physical'' coordinate takes on complex values, as discussed in Section \ref{sec:imag}.
Additionally, it is worth noting that the operators in the last term of the  \schr equation \eqref{eq:result} can be ordered differently. Namely, an alternative symmetrization leads to the \schr equation
\be
  \nonumber
  M^2 \varphi(z) =  -4 \tilde g^2\le({1\ov z}+{1\ov 1-z}\ri)\varphi(z) + [z(1-z)]^\half \, (i \p_z)^2 \, [z(1-z)]^\half \varphi(z)\, .
\ee
This equation is identical to the original \schr equation \eqref{eq:main} if we set
\be
  \nonumber
  \tilde g^2 = g^2 + {1\ov 16} \, .
\ee
Then, for  $g \in [0, {i\ov 4}) $, $\tilde g$ takes on real values in the range $\tilde g \in (0, {1\ov 4}]$. Since the AdS radius is stuck at $\mathcal{O}(\sqrt{\alpha'})$, there is no clear  notion of a classical target space, and it is our choice whether we interpret $g$ or $\tilde g$ as $ {R^2 \ov 2\pi \alpha'}$. In the latter case the radius of AdS is small, but positive.

The  Schrödinger equation was found to be equivalent to the \teller equation. By setting
\be
  \nonumber
  g={i }{\psyk-2 \ov 4\psyk} \, , \qquad \text{or equivalently:} \qquad \tilde g = {\sqrt{\psyk-1}\ov 2\psyk} \, ,
\ee
and applying a specific boundary condition to the antisymmetrized wavefunction (see equation \eqref{eq:bdyfin}), the resulting spectrum coincided with that of the disorder-averaged Sachdev-Ye-Kitaev model with $\psyk$-fermion interactions.

Why did we get this result? Perhaps it is due to the computational simplicity of the spectrum in equation \eqref{eq:codi}, with no deeper underlying connection to SYK. In fact, one of the motivating factors for exploring alternative string quantizations was to derive a spectrum simpler than that of the 't Hooft equation (appropriately generalized for AdS space \cite{Vegh:2023snc}). A more intriguing  possibility is that the bosonic string on a tiny AdS$_2$ target space, with radius comparable to the string scale, might be related to the bulk dual of SYK, as suggested in \cite{Maldacena:2016hyu}. According to the usual AdS/CFT dictionary, operator dimensions correspond to the masses of particles (or extended objects) in the bulk dual and we have shown that a simple folded string can indeed reproduce the spectrum of fermion bilinears. Whether this rigid model can be extended into a full-fledged dual theory is an interesting question for further investigation, which will be discussed in future work \cite{vegh1}.
It would be interesting to determine if the SYK spectrum computed in this paper is embedded within the full string theory with an arbitrary number of folds. Additionally, deriving the cubic couplings in the bulk by introducing an appropriate string interaction would be important, especially given that the corresponding fermion six-point functions have already been computed \cite{Gross:2017hcz, Gross:2017aos}.

The \teller system is well-known to be supersymmetric in the parameter region we considered in Section \ref{sec:imag}. It would be interesting to study the spectrum of the supersymmetric folded string. In order to make contact with the quantum spectral curve developed for strings in AdS$_5 \times S^5$, one needs to generalize our discussion to higher dimensions. In AdS$_d$ classical segmented strings have already been found, see \cite{Vegh:2015ska, Callebaut:2015fsa, Vegh:2016hwq, Gubser:2016wno, Gubser:2016zyw, Vegh:2016fcm, Vegh:2021jhl, Vegh:2021jqo, Vegh:2023snc}. Another potential direction is to place the string on a de Sitter target space, where the methods used in this paper could be similarly applied with $g \to -g$.  
Finally, we note that it may be possible to apply the quantum mechanical bootstrap programme \cite{Lin:2020mme, Han:2020bkb} to the folded string. The simplest case of $g=0$ and two folds has been addressed in \cite{Sword:2024gvv} where the bootstrap was found to produce exact results.

\vspace{0.2in}   \centerline{\bf{Acknowledgments}} \vspace{0.2in}
The author is grateful to Dionysios Anninos, Tarek Anous, Dami\' an Galante, Alessandro Georgoudis, and Lewis Sword for useful discussions, and Dionysios Anninos, Tarek Anous, and Dami\' an Galante for valuable comments on the draft.  
The author is supported by the STFC Consolidated Grant ST/T000686/1 ``Amplitudes, strings \& duality''. No new data were generated or analyzed during this study.

\bibliographystyle{JHEP}
\bibliography{paper}

\end{document}